%% file: paper.tex
\documentclass[twocolumn,showpacs,aps,prl,superscriptaddress]{revtex4}

\usepackage{xspace}
\usepackage{graphicx}
\usepackage{dcolumn}
\usepackage{amsmath}
\usepackage{epsfig}
\usepackage{multirow}

\input babarsym.tex

\newcommand{\taulg}      {\ensuremath{\mtau^{\pm} \to \ell^{\pm} \g}\xspace}
\newcommand{\taumg}      {\ensuremath{\mtau^{\pm} \to \mmu^{\pm} \g}\xspace}
\newcommand{\taueg}      {\ensuremath{\mtau^{\pm} \to e^{\pm} \g}\xspace}
\newcommand{\muelg}      {\ensuremath{\mmu^{\pm} \to \electron^{\pm} \g}\xspace}

\newcommand{\BRtaumg}    {\ensuremath{\BR(\taumg)}\xspace}
\newcommand{\BRtaueg}    {\ensuremath{\BR(\taueg)}\xspace}

\newcommand{\Mnu}              {\ensuremath{m_{\nu}^2}\xspace}

\def\recoilth {\ensuremath{\theta_{\mathrm{recoil}}}\xspace}
\def\recoilcth {\ensuremath{\cos \theta_{\mathrm{recoil}}}\xspace}
\def\cmcosopen {\ensuremath{\cos \theta_{\ell\g}}\xspace}

\def\DeltaEg {\ensuremath{\Delta E_\g}\xspace}
\def\DeltaE                    {\ensuremath{\Delta E}\xspace}

\newcommand{\eett}             {\ensuremath{e^+e^- \to \tautau}\xspace}

\def\eff                       {\ensuremath{\varepsilon}\xspace}

\def\teneight                  {\ensuremath{\times 10^{-8}}\xspace}

\newcommand{\eeeeg}   {\ensuremath{e^+e^- \to e^+e^-\gamma}\xspace}
\newcommand{\eemmg}   {\ensuremath{e^+e^- \to \mu^+\mu^-\gamma}\xspace}

\newcommand{\roots}        {\ensuremath{\sqrt{s}}\xspace}

\def\kk         {\mbox{\tt KK}\xspace}
\def\tauola     {\mbox{\tt TAUOLA}\xspace}
\def\photos     {\mbox{\tt PHOTOS}\xspace}

\def\evtgen     {\mbox{\tt EVTGEN}\xspace}
\def\jetset     {\mbox{\tt JETSET}\xspace}
\def\geant      {\mbox{\tt GEANT4}\xspace}

\newcommand{\gevccgevcc}{\ensuremath{{\mathrm{\,Ge\kern -0.1em V^2\!/}c^4}}\xspace}
\newcommand{\evcc}{\ensuremath{{\mathrm{\,e\kern -0.1em V\!/}c^2}}\xspace}
\newcommand{\CM} {\mbox{CM}\xspace}

\newcommand{\BABARPubYear}     {09}
\newcommand{\BABARPubNumber}  {026}
\newcommand{\SLACPubNumber} {13753}
\newcommand{\LANLNumber}  {0908.2381 [hep-ex]}

\def\figurebox#1#2#3{%
    \def\arg{#3}%
    \ifx\arg\empty
    {\hfill\vbox{\hsize#2\hrule\hbox to #2{\vrule\hfill\vbox to #1{\hsize#2\vfill}\vrule}\hrule}\hfill}%
    \else
    {\hfill\epsfbox{#3}\hfill}%
    \fi}

\begin{document}

\preprint{\babar-PUB-\BABARPubYear/\BABARPubNumber} 
\preprint{SLAC-PUB-\SLACPubNumber} 
\preprint{\LANLNumber}    

\begin{flushleft}
\babar-PUB-\BABARPubYear/\BABARPubNumber\\
SLAC-PUB-\SLACPubNumber\\
arXiv:\LANLNumber \\[10mm]    
\end{flushleft}

\title{{\large \bf \boldmath 
Searches for Lepton Flavor Violation in the Decays $\tau^\pm \rightarrow e^\pm \gamma$ and $\tau^\pm \rightarrow \mu^\pm \gamma$ 
}}

%
\input authors_jul2009_bad2218.tex

\date{\today}

\begin{abstract}
Searches for lepton-flavor-violating decays of a \mtau lepton to a lighter mass lepton and a photon
have been performed with the entire dataset of $(963 \pm 7) \times 10^6$ \mtau decays 
collected by the \babar\ detector near the \FourS, \ThreeS and \TwoS resonances.
The searches yield no evidence of signals and we set upper limits on the branching fractions of
\BRtaueg\ $<3.3\times10^{-8}$ and \BRtaumg\ $<4.4\times10^{-8}$ at 90\% confidence level.
\end{abstract}


\maketitle

Amongst all the possible lepton-flavor-violating \mtau processes,
\taulg (where $\ell = e, \mu)$ is predicted to be the dominant decay mode
in a wide variety of new physics scenarios, with rates close to current experimental limits.
Despite the existence of neutrino oscillations~\cite{NuOsc},
such decays are predicted to have unobservably low rates~\cite{Lee:1977ti} in the Standard Model (SM).
Thus, an observation of charged lepton flavor violation would be an unambiguous signature of new physics,
while improvements on existing limits will constrain many models.
As the relationships between \muelg, \taueg and \taumg decays are model-dependent,
searches for both \mtau modes provide independent information, 
even in the light of the small limit of $\BR(\mu^+\to\electron^+\gamma)<1.2\times10^{-11}$
at 90\% confidence level (C.L.)~\cite{Brooks:1999pu}.

Presently, the most stringent limits are \BRtaueg$<1.1\times10^{-7}$~\cite{Aubert:2005wa}
and \BRtaumg$<4.5\times10^{-8}$~\cite{Hayasaka:2007vc} at 90\% C.L.,
using 232.2\invfb and 535\invfb of \epem annihilation data 
collected near the \FourS resonance by the \babar\ and Belle experiments, respectively.
This paper reports the final result on these processes from BaBar.
It utilizes the entire dataset recorded by the \babar\ detector at the SLAC \pep2 \epem storage rings,
corresponding to a luminosity of
425.5 \invfb, 28.0 \invfb and 13.6 \invfb recorded at the \FourS, \ThreeS and \TwoS resonances,
and 44.4 \invfb,  2.6 \invfb and 1.4 \invfb recorded at 40\mev, 30\mev and 30\mev below the resonances, respectively.

For the bulk of the data sample at the \FourS resonance, 
the cross-section $\sigma_{\eett}$ = (0.919$\pm$0.003) nb~\cite{Banerjee:2007is},
determined to high precision using the \kk Monte Carlo (MC) generator~\cite{kk},
receives negligible contribution from \FourS due to its large decay width.
But, for the remaining data at the \ThreeS and \TwoS resonances, 
the \mtau-pair cross-section receives additional contributions of 
$\BR(\Upsilon\to\tau^+\tau^-) \approx 2\%$,
which are known only at the 10\% level~\cite{Amsler:2008zzb}.
Including a systematic uncertainty of 0.6\% on the luminosity determination,
this gives a total of $N_\tau = (963 \pm 7) \times 10^6$ \mtau decays.

The \babar\ detector is described elsewhere~\cite{detector}.
Charged particles are reconstructed as tracks with a 5 layer silicon vertex tracker
and a 40 layer drift chamber 
inside a 1.5 T solenoidal magnet. 
A CsI(Tl) electromagnetic calorimeter 
is used to identify electrons and photons. A ring-imaging Cherenkov detector 
is used to identify charged pions and kaons. The flux return of the solenoid,  
instrumented with resistive plate chambers 
and limited streamer tubes, is used to identify muons.


The signal is characterized by a $\ell^\pm\gamma$ pair 
with an invariant mass and total energy in the center-of-mass (\CM) frame ($E^{\mathrm{CM}}_{\ell\g}$)
close to $m_\tau$ = 1.777\gevcc~\cite{Amsler:2008zzb} and \roots/2, respectively.
The event must also contain another \mtau decay, reconstructed as decaying to one or three tracks.

The dominant irreducible background comes from \mtau-pair events containing hard photon radiation
and one of the \mtau leptons decaying to a charged lepton.
The remaining backgrounds for \taueg and \taumg decays arise from the relevant radiative processes, 
\eeeeg and \eemmg, and from hadronic \mtau decays where a pion is misidentified as the electron or muon.

Signal events are simulated using \kk and \tauola~\cite{tauola} with measured \mtau branching fractions~\cite{Amsler:2008zzb}.
The $\mu^+\mu^-$ and \tautau background processes are generated using \kk and \tauola,
while the \qqbar processes are generated using \jetset~\cite{Sjostrand:1995iq} and \evtgen~\cite{Lange:2001uf}.
Radiative corrections for all processes are simulated using \photos~\cite{Golonka:2005pn}.
The Bhabha background is studied using events with two identified electrons in the data.
The two-photon background has been studied and found to be negligible.
The detector response to generated particles is simulated using the \geant package~\cite{geant}.
MC events are used to optimize the selection criteria and estimate the systematic uncertainties on the efficiency,
while the background rates are estimated directly from data.

Events with two or four well reconstructed tracks and zero total charge are selected,
where no track pair is consistent with being a photon conversion in the detector material.
Each event is divided into hemispheres (``signal-'' and ``tag-'' sides) in the \CM\ frame 
by a plane perpendicular to the thrust axis
calculated using all reconstructed charged and neutral particles~\cite{thrust}.

The signal-side hemisphere must contain one photon with \CM\ energy $E^{\mathrm{CM}}_\gamma$ greater than 1 \gev,
and no other photon with energy greater than 100 \mev in the laboratory frame.
The signal side must contain one track within the calorimeter acceptance 
with momentum in the \CM\ frame less than $0.77\roots/2$.
This track must be identified as an electron or a muon for the \taueg or \taumg search. 
The electron selectors have an efficiency of 96\% within the fiducial coverage.
For reliable muon identification, the track momentum is required to be greater than 0.7 \gevc in the laboratory frame,
above which the selection efficiency is 83\%.

In the rest-frame of the $\mtau^\pm$, the $\ell^\pm$ and the $\g$ are produced back-to-back.
When boosted to the \CM\ frame, kinematic considerations of two-body decays require there to be a minimum opening angle between them.
The cosine of the opening angle, \cmcosopen, between signal-track and signal-photon is required to be less than 0.786.

The tag-side hemisphere is expected to contain a SM \mtau decay.
A tag-side hemisphere containing a single track is classified as $e$-tag,  $\mu$-tag, or  $\pi$-tag 
if the total photon \CM\ energy in the hemisphere is less than 200\mev and 
the track is exclusively identified as an electron ($e$-tag), as a muon ($\mu$-tag), or as neither ($\pi$-tag). 
Events with the tag-side track failing both the lepton selectors are classified as $\rho$-tag
if they contain at least one \piz candidate reconstructed from a pair of photons with invariant mass between 90 and 165 \mevcc.
If the tag-side hemisphere contains three charged tracks, all of which fail the lepton identification, it is classified as a $3h$-tag.

The definitions of the tag-side modes are designed to minimize the residual backgrounds 
from radiative QED processes.
For the \taueg search, very loose electron selection criteria are applied for the $e$-tag sample.
Thus, the remaining tags which fail these very loose electron criteria have small Bhabha contamination.
The $e$-tag events are used as the control sample to model the Bhabha background characteristics,
and are removed from the final sample of events in the \taueg search.
Similarly, for the \taumg search, very loose muon criteria are applied for the $\mu$-tag, on which stricter kinematic requirements are later imposed
with tolerable loss in signal efficiency. The other tags are required to fail these very loose muon criteria, thereby reducing di-muon backgrounds.

To suppress non-\mtau backgrounds with missing momentum along the beam direction due to initial and final state photon radiation,
we require that the polar angle $\theta_{\rm{miss}}$ of the missing momentum 
be inside the detector acceptance, i.e. $-0.76 < \cos\theta_{\rm{miss}} < 0.92$.

The total \CM\ momentum of all tracks and photon candidates on the tag-side
is required to be less than $0.77\roots/2$ for $e$-, $\mu$-, $\pi$-tags and less than $0.9\roots/2$ for $\rho$- and $3h$-tags.
The tag-side pseudomass~\cite{Albrecht:1992td} is required to be less than 0.5\gevcc for $e$-, $\mu$-, $\pi$-tags and 
less than 1.777\gevcc for $\rho$- and $3h$-tags.

The mass squared \Mnu of the missing particles on the tag side is calculated
using the tag-side tracks and photon candidates and assuming that in the \CM\ frame, 
the tag-side \mtau momentum is opposite that of the signal \mtau and that its energy is \roots/2.
To reduce backgrounds, we require $\Mnu > -0.25 \gevccgevcc$ for $e$- and $\mu$-tags,
$|\Mnu| < 0.25 \gevccgevcc$ for $\pi$- and $3h$-tags, and
$|\Mnu| < 0.50 \gevccgevcc$ for $\rho$-tags.

For radiative Bhabha and di-muon events, 
the expected photon energy in the \CM frame 
$(E_{\gamma}^{\mathrm{CM}})_{\mathrm{exp}}$ is 
$\frac{|\sin(\theta_1+\theta_2)|\roots}{\sin\theta_1+\sin\theta_2+|\sin(\theta_1+\theta_2)|}$,
where $\pi-\theta_1$ and $\pi-\theta_2$ are the angles the photon momentum 
makes with the signal-track and the total observed tag-side momentum, respectively.
Also, for such events, we expect the cosine of the opening angle, \recoilth, 
between the signal-track and the total observed tag-side momentum
in the reference frame obtained by removing the signal photon from the CM frame
to peak at -1. To suppress these backgrounds, 
we remove events having reconstructed photon energy consistent with the expected value, i.e.
$|{E^{\mathrm{CM}}_\gamma} - (E_{\gamma}^{\mathrm{CM}})_{\mathrm{exp}}| \equiv |\DeltaEg| < 0.125\roots$
and  $\recoilcth < -0.975$ in $e$- and $\mu$-tags for the \taumg search.
No such criteria are necessary for the \taueg search according to the optimization procedure.

To further suppress the remaining backgrounds, neural net (NN) based discriminators 
are employed separately for each tag and 
for each dataset taken at values of $\sqrt{s}$ 
near the \FourS, \ThreeS and \TwoS resonances.
Six observables are used as input to the NN:
the total tag-side momentum divided by \roots/2,
\Mnu, \DeltaEg/\roots, \recoilcth, \cmcosopen, and 
the transverse component of missing momentum relative to the collision axis.
The NN based discriminators improve the signal to background ratios for the two searches
by factors of 1.4 and 1.3, respectively.

Signal decays are identified by two kinematic variables: 
the energy difference $\DeltaE = E^{\mathrm{CM}}_{\ell\g} - \roots/2$ and 
the beam-energy constrained \mtau mass (\mec), obtained from a kinematic fit after requiring the \CM\ \mtau energy to be \roots/2
and after assigning the origin of the \g candidate to the point of closest approach of the signal lepton track to the \epem collision axis.
The distributions of these two variables have a small correlation arising from initial- and final- state radiation.
For signal MC events, the \mec and \DeltaE distributions are centered at $m_\tau$ and small negative values, respectively,
where the shifts from zero for the latter are due to radiation and photon energy reconstruction effects.
The mean and standard deviations of the \mec and \DeltaE distributions 
for the reconstructed signal MC events are presented in Table~\ref{tab:table1}. 
The data events falling within a $3\sigma$ ellipse in the \mec vs. \DeltaE plane,
centered around the reconstructed peak positions as obtained using signal MC,
are not examined until all optimization and systematic studies have been completed.
The selections are optimized to yield the smallest expected upper limits~\cite{Feldman:1997qc}
for observing events inside a $2\sigma$ signal ellipse 
under background-only hypotheses.

\renewcommand{\multirowsetup}{\centering}
\newlength{\LL}\settowidth{\LL}{$8047$}
\begin{table*}[!htbp]
\begin{center}
\caption{Means and resolutions of \mec and \DeltaE distributions for the signal MC events,
the numbers of observed (obs) and expected (exp) events inside the $2\sigma$ signal ellipse,
the signal efficiencies (\eff), and the 90\% C.L. upper limits (UL).}
\begin{ruledtabular}
\renewcommand{\arraystretch}{1.10}
{\scalebox{0.99}{
\begin{tabular}{l|c|c|c|c|c|c|c|c|c}
Decay modes  &$<\mec>$&$\sigma(\mec)$&$<\DeltaE>$&$\sigma(\DeltaE)$ &\multicolumn{2}{c|}{$2\sigma$ signal ellipse} &\eff&\multicolumn{2}{c}{UL (\teneight)}\\
\cline{2-10}
             &   \mevcc  & \mevcc       & \mev    &\mev              &obs&exp                             &(\%)&obs&exp  \\\hline
\taueg   &1777.3 &8.6 &-21.4& 42.1 & 0 & 1.6$\pm$0.4 & 3.9$\pm$0.3 & 3.3  & 9.8   \\\hline\hline
\taumg   &1777.4 &8.3 &-18.3& 42.2 & 2 & 3.6$\pm$0.7 & 6.1$\pm$0.5 & 4.4  & 8.2   \\
\end{tabular}
}}
\end{ruledtabular}
\label{tab:table1}
\end{center}
\end{table*}

\renewcommand{\multirowsetup}{\centering}

\begin{figure*}[!htbp]
\vskip 0.8cm
\resizebox{\textwidth}{.27\textheight}{%
\includegraphics{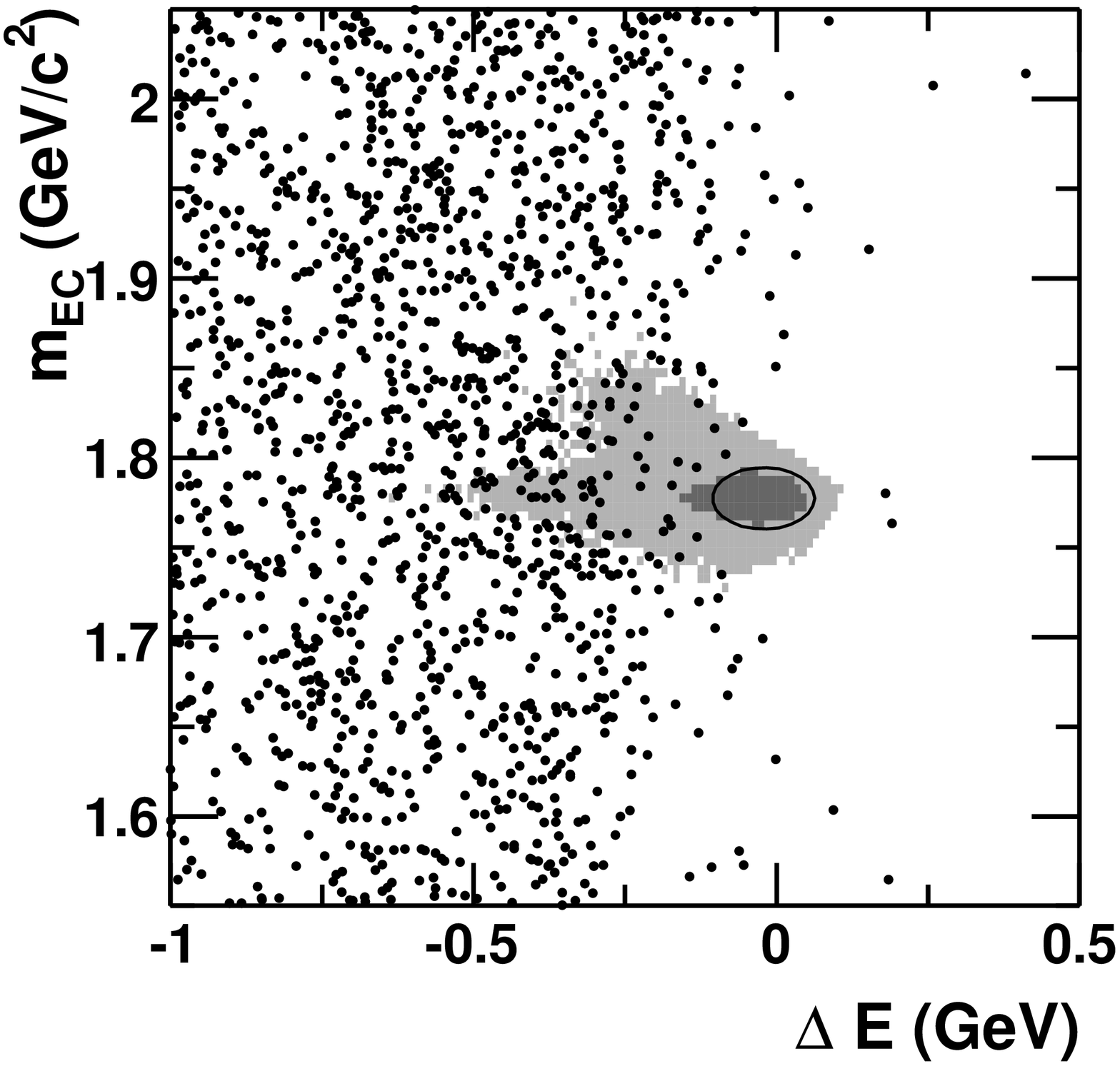}
\hskip 2.0cm
\includegraphics{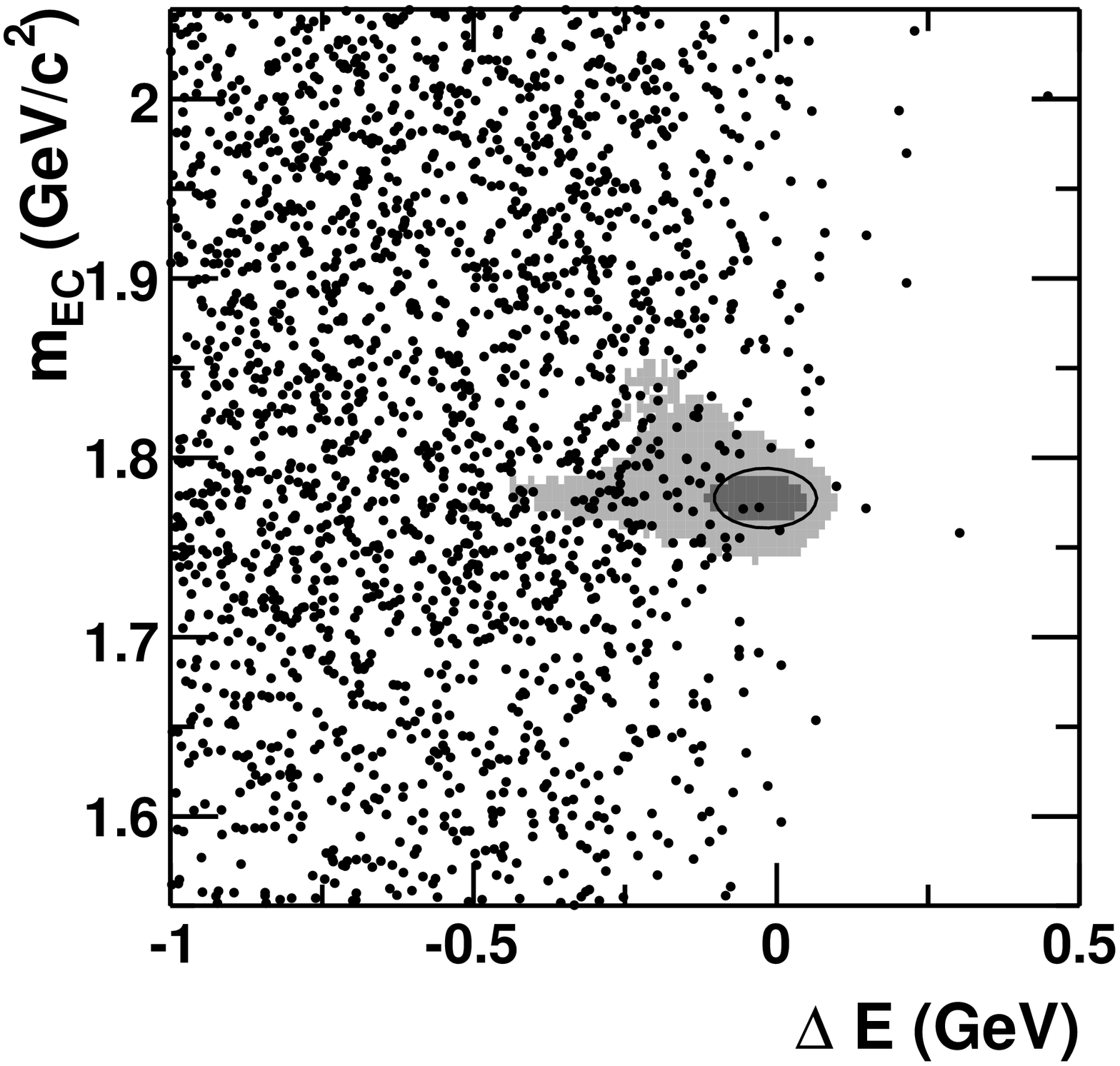}
}
\caption{The GSB and the $2\sigma$ ellipse for \taueg (left) and \taumg (right) decays in the \mec vs. \DeltaE plane.
         Data are shown as dots and contours containing 90\% (50\%) of signal MC events
         are shown as light- (dark-) shaded regions.}
\label{fig1}
\end{figure*}

The distributions of events in \mec vs. \DeltaE are shown in Fig.~\ref{fig1}. 
To study signal-like events, a Grand Signal Box (GSB) is defined as
\mec $\in$ $[1.55, 2.05]$ \gevcc and \DeltaE $\in$ $[-1.0, 0.5]$ \gev.
Outside the blinded $3\sigma$ ellipse, 1389 data events survive in the GSB for the \taueg channel,
and 2053 data events survive for the \taumg channel.
These agree to within 2.4\% and 1.7\% with the numbers of background MC events observed.
The signal-track arises from a real electron or muon in 96\% and 82\% of the background MC events for the two searches.

A Fit Box (FB) region is defined as \mec $\in$ $[1.6, 2.0]$ \gevcc and \DeltaE $\in$ $[-0.14, 0.14]$ \gev, 
excluding the blinded $3\sigma$ ellipse.
The \mec vs. \DeltaE distributions of events inside the FB are modeled 
by 2-dimensional probability density functions (PDFs) summed over all background event types.
The PDFs have correlations built in using Gaussian weights with an adaptive kernel estimation procedure~\cite{keys}.
The shape of the Bhabha component is obtained using the data samples 
having $\recoilcth < -0.8$ from events selected in the $e$-tag sample for the \taueg search,
while the shapes of \mumu, \tautau and \qqbar PDFs are obtained from their respective MC samples.

The fractions of events for each background type are obtained from separate maximum likelihood fits 
to 41 and 105 events inside the FB, respectively, for the two searches.
We find $(70 \pm 15)$\% and $(90 \pm 8)$\% of the background events are \mtau-pair events.
By integrating the total PDF summed over background types only, 
we expect $(1.6\pm0.3)$ and $(3.6\pm0.4)$ events
inside the $2\sigma$ signal ellipse for the two searches,
where the quoted statistical errors are due to the sizes of the fitted dataset.

As a cross-check, we integrate the total PDF over four $2\sigma$ ellipses inside the FB, whose centers
are shifted by $\pm5\sigma$ or by $\pm9\sigma$ along \mec only.
The numbers of observed events in each of these neighboring regions and their sums
are consistent with the expected numbers of events, 
which are shown along with their statistical errors in Table~\ref{tab:table2}.

\begin{table}[!h]
\begin{center}
\caption{Numbers of observed (obs) and expected (exp) numbers of background events along with statistical errors 
inside $2\sigma$ ellipses whose centers are shifted by $\pm 5\sigma$ and $\pm 9 \sigma$ in \mec only, and their sums.}
\begin{tabular}{l|c|c|c|c|c|c} \hline
            &      & -9$\sigma$    & -5 $\sigma$  & +5 $\sigma$   & +9 $\sigma$ & sum\\ \hline 
\taueg      & obs  &  2            &    1         & 2             & 2           & 7  \\ \cline{2-7}
            & exp  &  1.2$\pm$0.2  & 1.4$\pm$0.2  & 1.9$\pm$0.3   & 2.1$\pm$0.3 & 6.6$\pm$0.5 \\ \hline
\taumg      & obs  &  3            &    1         & 4             & 6           & 14 \\ \cline{2-7}
            & exp  &  2.8$\pm$0.3  & 3.1$\pm$0.3  & 4.2$\pm$0.4   & 4.8$\pm$0.5 & 14.9$\pm$0.8\\ \hline
\end{tabular}
\label{tab:table2}
\end{center}
\end{table}

To obtain the systematic errors on the numbers of expected background events, 
we fit the \mec distributions of 32 and 81 data events inside the $\pm2\sigma$ band in \DeltaE 
over the GSB region but outside the blinded $3\sigma$ ellipse.
Varying degrees of polynomial functions are used to model the \mec distributions,
which are then integrated to obtain the number of expected events inside the $2\sigma$ ellipse.
The largest deviations between the predictions from 1-dimensional and 2-dimensional fits
are used to set the total uncertainties of 0.4 and 0.7 events on the background estimates.

The systematic uncertainties in the signal selection and reconstruction efficiencies for \taueg and \taumg decays
due to the modeling of the variables entering the NN are 2.7\% and 1.8\%, respectively.
Those due to the photon reconstruction efficiency are 1.8\% for both decays,
while those due to the signal-lepton track identification are 2.3\% and 2.7\%, respectively.
The contributions due to the uncertainty in the signal-track momentum and signal-photon energy scale and resolution,
estimated by varying the peak position and resolution of the \mec and \DeltaE distributions, are 6.4\%  and 6.2\%, respectively. 
Other systematic uncertainties totaling less than 1.5\% for both signal decay modes
include those arising from trigger and filter efficiencies, tracking efficiencies, and the beam-energy scale and spread. 
We use approximately $10^6$ MC events per channel, resulting in a negligible systematic uncertainty due to MC statistics.
Although the signal MC has been modeled using a flat phase space model, the efficiencies 
are insensitive to this assumption as demonstrated by considering the two 
extreme cases of $V-A$ and $V+A$ forms of interaction for the signal MC. 
All contributions to the systematic uncertainties are added in quadrature to give total relative 
systematic uncertainties on the efficiencies of 7.7\% and 7.4\% for \taueg and \taumg decays, respectively. 

We observe 0 and 2 events for the \taueg and \taumg searches inside the 2$\sigma$ signal ellipse, respectively.
As there is no evidence for a signal, we set a frequentist upper limit 
calculated using $\BR^{90}_{UL}=N^{90}_{UL}/(N_{\tau}\eff)$ to be 
\BRtaueg$<$ 3.3 $\times$ 10$^{-8}$ and \BRtaumg$<$ 4.4 $\times$ 10$^{-8}$ at 90\% C.L.,
where \eff is the signal efficiency inside the 2$\sigma$ signal ellipse
and $N^{90}_{UL}$ is the 90\% C.L. upper limit on the number of signal events,
estimated using the POLE program~\cite{Conrad:2002kn}.
The upper limits which include all systematic uncertainties,
are presented in Table~\ref{tab:table1}, along with signal efficiencies and numbers of observed and expected background events.
These results supersede previous \babar\ results~\cite{Aubert:2005wa, Aubert:2005ye},
reducing the upper limits by factors of 3.3 and 1.5, respectively, 
and are the most stringent limits on searches for lepton flavor violation in \taueg and \taumg decays.

\input acknow_PRL.tex


\end{document}

%% file: authors_jul2009_bad2218.tex
%
\author{B.~Aubert}
\author{Y.~Karyotakis}
\author{J.~P.~Lees}
\author{V.~Poireau}
\author{E.~Prencipe}
\author{X.~Prudent}
\author{V.~Tisserand}
\affiliation{Laboratoire d'Annecy-le-Vieux de Physique des Particules (LAPP), Universit\'e de Savoie, CNRS/IN2P3,  F-74941 Annecy-Le-Vieux, France}
\author{J.~Garra~Tico}
\author{E.~Grauges}
\affiliation{Universitat de Barcelona, Facultat de Fisica, Departament ECM, E-08028 Barcelona, Spain }
\author{M.~Martinelli$^{ab}$}
\author{A.~Palano$^{ab}$ }
\author{M.~Pappagallo$^{ab}$ }
\affiliation{INFN Sezione di Bari$^{a}$; Dipartimento di Fisica, Universit\`a di Bari$^{b}$, I-70126 Bari, Italy }
% ...
\collaboration{The \babar\ Collaboration}
\noaffiliation

%% file: acknow_PRL.tex
We are grateful for the excellent luminosity and machine conditions
provided by our \pep2\ colleagues, 
and for the substantial dedicated effort from
the computing organizations that support \babar.
The collaborating institutions wish to thank 
SLAC for its support and kind hospitality. 
This work is supported by
DOE
and NSF (USA),
NSERC (Canada),
CEA and
CNRS-IN2P3
(France),
BMBF and DFG
(Germany),
INFN (Italy),
FOM (The Netherlands),
NFR (Norway),
MES (Russia),
MEC (Spain), and
STFC (United Kingdom). 
Individuals have received support from the
Marie Curie EIF (European Union) and
the A.~P.~Sloan Foundation.